%% file: adbis.tex
\documentstyle[runningheads]{llncs}

\begin{document}

\pagestyle{headings}

\input{macros}

\input{paper-macros}
\mainmatter

\input{title}

\newpage

\input{abstract}

\input{intro}        % 640  words
\input{motivation}   % 500  words
\input{framework}    % 1546 words
\input{rewritings}   % 451  words
\input{count_views}  % 4000 words (w.o. max 3494) -> 2951
\input{conclusion}   % 155  words
\bibliography{%/home/nutt/Literature/strings,%
	      %/home/nutt/Literature/literature,%
	      %strings,%
              %literature,%
	      %/cs/visitor/nutt/Literature/lit-david,%
	      adbis} %lit-sara}%,
\bibliographystyle{alpha}
\newpage
\input{appendix}

\end{document}

%% file: macros.tex
%%%%%%%%%%%%%%%%%%%%%%%%%%%%%%%%%%%%%%%%%%%%% General Math
\def\A{{\cal A}} \def\B{{\cal B}} \def\C{{\cal C}} \def\D{{\cal D}}
\def\E{{\cal E}} \def\F{{\cal F}} \def\G{{\cal G}} \def\H{{\cal H}}
\def\I{{\cal I}} \def\J{{\cal J}} \def\K{{\cal K}} \def\L{{\cal L}}
\def\M{{\cal M}} \def\N{{\cal N}} \def\O{{\cal O}} \def\P{{\cal P}}
\def\Q{{\cal Q}} \def\R{{\cal R}} \def\S{{\cal S}} \def\T{{\cal T}}
\def\U{{\cal U}} \def\V{{\cal V}} \def\W{{\cal W}} \def\X{{\cal X}}
\def\Y{{\cal Y}} \def\Z{{\cal Z}}

\newcommand{\dd}[2]{#1_1,\ldots,#1_{#2}}      % da da, makes x1,...,xn

\newcommand{\set}[1]{\{#1\}}
\newcommand{\eset}{\emptyset}
\newcommand\bigset[1]{ \Bigl\{ #1 \Bigr\} }   % makes the big set { #1 }
\newcommand\bigmid{\ \Big|\ }

\newcommand{\incl}{\subseteq}		% included
\newcommand{\incls}{\supseteq}		% includes

\newcommand{\col}{\colon}

\newcommand{\NP}{{\rm NP}}		% the complexity class NP
\newcommand{\GI}{{\rm GI}}              % the complexity class GI
\newcommand{\PIPEETWO}{\Pi^{{\rm P}}_2}	% the complexity class PiPeeTwo
\newcommand{\SIGPEETWO}{\Sigma^{{\rm P}}_2}	
					% the complexity class SigmaPeeTwo
\newcommand{\PSPACE}{{\rm PSPACE}}	% complexity class PSPACE
\newcommand{\PTIME}{{\rm P}}		% complexity class PTIME
\newcommand{\EXPTIME}{{\rm EXPTIME}}	% deterministic exponential time

\newcommand{\angles}[1]{\langle#1\rangle}	% pointed angles

%%%%%%%%%%%%%%%%%%%%%%%%%%%%%%%%%%%%%%%%%%%%% Proofs

\newcommand{\OnlyIf}{\lq\lq$\Rightarrow$\rq\rq\ \ }   % 1st direction of iff-proof
\newcommand{\If}{\lq\lq$\Leftarrow$\rq\rq\ \ }        % 2nd direction of iff-proof

%%%%%%%%%%%%%%%%%%%%%%%%%%%%%%%%%%%%%%%%%%%%% Text Processing
\newcommand{\quotes}[1]{\lq\lq#1\rq\rq}     	% makes ``#1''
\newcommand{\wrt}{w.r.t.}			% with respect to 
\newcommand{\WLOG}{w.l.o.g.}			% without loss of generality
\newcommand{\ie}{i.e.}		        	% i.e.
\newcommand{\eg}{e.g.}		        	% i.e.
\renewcommand{\hom}{homo\-mor\-phism}

\newcommand{\eat}[1]{}

\newcommand{\comment}[1]{\noindent{\sl COMMENT:} {\sl #1}}

%% file: paper-macros.tex
\newcommand{\per}{{\bf .}}        	% period
\newcommand\osp{\;}                     % operator space,
					% put before and after operators
\newcommand\osps[1]{{\osp{#1}\osp}}     % operator spaces

\newcommand{\id}{{\sl id}}		% the identity mapping

%%%%%%%%%%%%%%%%%%%%%%%%%%%%% Syntax of queries

\newcommand{\tpl}[1]{\bar{#1}}	        % tuple
\newcommand{\ar}[2]{#1^{(#2)}}		% arity of a predicate
\newcommand{\SIG}{\Sigma}		% signature

\newcommand{\type}{\tau}  		% generic symbol for a data type

\newcommand\AND{\osps{\&}}              % and in a query
\newcommand\OR{\vee}                    % or in a query
\newcommand\BIGOR{\bigvee}              % big or in a query
\newcommand{\qif}{\leftarrow}           % query if

\newcommand{\having}{{\tt having}}      % "having in SQL"
\newcommand{\union}{{\tt union}}        % "union in SQL"

%%%%%%%%%%%%%%%%%%%%%%%%%%%%% Semantics of queries

\newcommand{\carr}[1]{|#1|}             % carrier of a database

\newcommand{\DB}[2]{#1^{#2}}		% database relation
\newcommand{\DBD}[1]{\DB{#1}{\D}}	% database relation over db D
\newcommand{\DBE}[1]{\DB{#1}{\E}}	% database relation over db E

\newcommand{\DBN}{\D_{\tpl N}}		% family of databases (``expansions'')
					% constructed from 

\newcommand{\bag}[1]{\{\!\!\{#1\}\!\!\}}    
                                        % bag, multiset

%\newcommand{\DBBAG}[2]{{#1}_b^{#2}}	% bag resulting from a query over \D
\newcommand{\DBBAG}[2]{\bag{#1}^{#2}}	% bag resulting from a query over \D
\newcommand{\DBDBAG}[1]{\DBBAG{#1}{\D}}	% bag resulting from a query over \D

\newcommand{\MULT}[1]{\M(#1)}		% set of multisets over #1

\newcommand{\bigbag}[1]{\Bigl\{\!\!\Bigl\{#1\Bigr\}\!\!\Bigr\}}    
                                        % big bag, big multiset

%%%%%%%%%%%%%%%%%%%%%%%%%%%%% Aggregation functions
\newcommand{\MIN}{{\sl min}}      	% minimum
\newcommand{\MAX}{{\sl max}}      	% maximum
\newcommand{\COUNT}{{\sl count}}  	% count
\newcommand{\CNTD}{{\sl cntd}}  	% count distinct
\newcommand{\SUM}{{\sl sum}}      	% sum
\newcommand{\AVG}{{\sl avg}}      	% average

%%%%%%%%%%%%%%%%%%%%%%%%%%%%% Semantics of aggregate queries
\newcommand{\CLASS}[2]{[#1]_{#2}}      	% equivalence class 

\newcommand{\eqvq}{\sim_{q}} 		% equivalence of assignments
\newcommand{\eqv}{\sim} 		% equivalence of assignments
\newcommand{\classq}[1]{\CLASS{#1}{q}} 	% equivalence class of an assignment wrt q
\newcommand{\class}[1]{[#1]}        	% equivalence class of an assignment w/o query

\newcommand{\deqv}{\approx} 		% double equivalence sign
\newcommand{\dclass}[1]{\CLASS{#1}{\deqv}}	
					% equivalence class wrt \deqv

\newcommand{\card}[1]{|#1|}        	% cardinality of a set
\newcommand{\bigcard}[1]{\Bigl|#1\Bigr|}        	
					% cardinality of a set
\newcommand{\CARD}[1]{{\sl card}#1}    	% cardinality of a set

\newcommand{\Ker}[2]{#1_{#2}}        	% Ker q j  is the j-th kernel 
					% of query q

%\newcommand{\core}[1]{\hat#1}        	% core of an aggregate query
\newcommand{\core}[1]{\breve{#1}}        	% core of an aggregate query
\newcommand{\extinst}[1]{\hat{#1}}      % extended instantiation

%%%%%%%%%%%%%%%%%%%%%%%%%%%%% Orderings

\newcommand{\pod}{\preceq}              % preorder
\newcommand{\dop}{\succeq}              % converse of a preorder
\newcommand{\spod}{\prec}               % strict preorder
\newcommand{\sdop}{\succ}               % strict converse of a preorder

\newcommand{\phiuv}{\phi_{\tpl u}^{\tpl v}}
					% monotonic interpolation function

%%%%%%%%%%%%%%%%%%%%%%%%%%%%% Numbers
\newcommand{\rat}{{\bf Q}}		% the rational numbers
\newcommand{\nat}{{\bf N}}		% the natural numbers
\newcommand{\intg}{{\bf Z}}		% the integers
\newcommand{\real}{{\bf R}}		% the real numbers

%%%%%%%%%%%%%%%%%%%%%%%%%%%%% Orderings

\newcommand{\qcount}{\Gamma}		% counting function over \DBN

\newcommand{\HOM}{{\sl Hom}}

\newcommand{\ptrn}{\P}

\newcommand{\comp}[1]{\tilde #1}	% completion of a set of comparisons

\newcommand{\essnovar}[1]{{\sharp}_{\sl v}(#1)}
					% essential number of variables in 
					% a query
\newcommand{\essnorel}[1]{{\sharp}_{\sl r}(#1)}
					% essential number of variables in 
					% a query

\newcommand{\red}[1]{#1^{{\sl r}}}	% reduced version of a set of 
					% comparisons
\newcommand{\eq}[1]{#1^{=}}		% set of equalities

\newcommand{\intmodels}{\models_\intg}	% implication over the integers
\newcommand{\ratmodels}{\models_\rat}	% implication over the rationals

\newcommand{\udist}[3]{{\sl dist}^{\uparrow}_{#1}(#2,#3)}
					% upward distance from a variable
					% to a constant
\newcommand{\ddist}[3]{{\sl dist}^{\downarrow}_{#1}(#2,#3)}
					% downward distance from a variable
					% to a constant
\newcommand{\udistc}[2]{\udist{C}{#1}{#2}}
					% upward distance from a variable
					% to a constant wrt C
\newcommand{\ddistc}[2]{\ddist{C}{#1}{#2}}
					% downward distance from a variable
					% to a constant wrt C

\newcommand{\GLL}{{\sl gll}}
\newcommand{\GLLC}{{\sl gll}_C}
\newcommand{\gll}[2]{\GLL_{#1}({#2})}
					% greatest lower limit of a variable
					% wrt to a set of comparisons
\newcommand{\LUL}{{\sl lul}}
\newcommand{\LULC}{{\sl lul}_C}
\newcommand{\lul}[2]{\LUL_{#1}(#2)}
					% least upper limit of a variable
					% wrt to a set of comparisons
\newcommand{\gllc}[1]{\gll{C}{#1}}
					% greatest lower limit of a variable
					% wrt to a set of comparisons
\newcommand{\lulc}[1]{\lul{C}{#1}}
					% least upper limit of a variable
					% wrt to a set of comparisons
\newcommand{\glll}[1]{\gll{L}{#1}}
					% greatest lower limit of a variable
					% wrt to a set of comparisons L
\newcommand{\lull}[1]{\lul{L}{#1}}
					% least upper limit of a variable
					% wrt to a set of comparisons L

\newcommand{\gamdown}{\gamma^{\downarrow}}
					% lower limit assignment
\newcommand{\gamup}{\gamma^{\uparrow}}
					% uper limit assignment

\newcommand{\YPLUS}{Y^{+}}
\newcommand{\YMIN}{Y^{-}}
\newcommand{\dmin}{d^{-}}
\newcommand{\dplus}{d^{+}}

\newcommand{\VIRT}[1]{{\sl vc}_{#1}}	% virtual constants of #1
\newcommand{\VIRTC}{\VIRT C}		% virtual constants of C
\newcommand{\virt}[2]{{\sl vc}_{#1}(#2)}% virtual constants of #1 wrt #2
\newcommand{\virtc}[1]{\virt{C}{#1}} 	% virtual constants of C wrt #1

%%%%%%%%%%%%%%%%%%%%%%%%%% Linearizations

\newcommand{\lin}[2]{\L_{#1}(#2)}	% linearizations of the terms in
					% a query
\newcommand{\lind}[1]{\lin{D}{#1}}	% linearizations for set of constants D

\newcommand{\varl}[2]{\L^{\sl v}_{#1}(#2)}	% linearizations where summation term 
					% is a variable
\newcommand{\varld}[1]{\varl{D}{#1}}	% linearizations for set of constants D

\newcommand{\consl}[2]{\C\L_{#1}(#2)}	% linearizations where summation term 
					% is a constant
\newcommand{\consld}[1]{\consl{D}{#1}}	% linearizations for set of constants D

\newcommand{\ndcqd}{N_D({\core q},d)}   % number of linearizations of core q with y = d
\newcommand{\ndcqpd}{N_D({\core q'},d)} % number of linearizations of core q' with y = d

\newcommand{\q}[1]{q_{#1}}		% linearization of q wrt #1
\newcommand{\ql}{\q{L}}			% linearization of q wrt L
\newcommand{\qll}{(\ql)_L}		% linear expansion of q
\newcommand{\qp}[1]{q'_{#1}}		% linearization of q' wrt #1
\newcommand{\qpm}{\qp{M}}		% linearization of q' wrt M
\newcommand{\qpmm}{(\qpm)_{M}}		% linear expansion of q'

\newcommand{\cq}[1]{\core{q}_{#1}}	% linearization of core q wrt #1
\newcommand{\cql}{\cq{L}}		% linearization of core q wrt L
\newcommand{\cqil}{\cq{iL}}		% linearization of core q wrt L
\newcommand{\cqll}{(\cql)_L}		% linear expansion of core q
\newcommand{\cqp}[1]{\core{q}'_{#1}}	% linearization of core q' wrt #1
\newcommand{\cqpm}{\cqp{M}}		% linearization of core q' wrt M
\newcommand{\cqpjm}{\cqp{jM}}		% linearization of core q' wrt M
\newcommand{\cqpmm}{(\cqpm)_{M}}	% linear expansion of core q'

\newcommand{\wiso}{\sim}		% weak isomorphims of classes of queries

\newcommand{\phil}{\phi_L}		% canonical mapping for L
\newcommand{\phim}{\phi_M}		% canonical mapping for M

\newcommand{\sig}[1]{\sigma(#1)}	% selects the summation constant

\newcommand{\QV}{\Q_\V}			% classes of variable queries
\newcommand{\QC}{\Q_\C}			% classes of constant queries

\newcommand{\rel}[1]{{#1}^{\sl rel}}	% relational part of a query

\newcommand{\lnkd}{\leftrightarrow}	% ``linked,'' an equivalence 
					% relation on terms

%%%%%%%%%%%%%%%%%%%%%%%%%%%%%%%%%%%%

\newcommand{\lexp}[1]{{#1}^{{\sl lin}}}	% linear expansion of a query

\newcommand{\expqll}{\BIGOR_L \ql}	% linear expansion of q
\newcommand{\expqpmm}{\BIGOR_M \qm}	% linear expansion of q'

%%%%%%%%%%%%%%%%%%%%%%%%%%%%%%%%%%%%    Miscelanneous

\newcommand{\inv}[1]{#1^{-1}}           % inverse of a mapping etc.

\newcommand{\restr}[2]{#1_{|#2}}        % inverse of a mapping etc.

\newcommand{\var}[1]{{\sl var}(#1)}     % variables in a syntactic entity

\newcommand{\NDVAR}[1]{{\sl ndv}(#1)}	% number of nondistinguished variables in a query

%%%%%%%%%%%%%%%%%%%%%%%%%%%%%%%%%%%  Rewriting Using Views

\newcommand{\EXTDB}[2]{#1_#2}		% extension of db #1 by views defined
					% through #2

\newcommand{\equivv}{\equiv_\V}		% equivalence wrt set of views \V
\newcommand{\equivvp}{\equiv_{\V'}}	% equivalence wrt set of views \V

\renewcommand{\exp}[1]{\tilde #1}	% unfolding wrt set of views \V

\newcommand{\unf}[1]{#1^{{\sl u}}}	% unfolding wrt a set of views 

\newcommand{\Count}{{\sl c}}			% superscript ``count''
\newcommand{\Sum}{{\sl s}}			% superscript ``sum''
\newcommand{\Max}{{\sl m}}			% superscript ``max''
\newcommand{\Reg}{{\sl r}}			% superscript ``regular''

\newcommand{\VC}{\V^\Count}			% set of count views
\newcommand{\VS}{\V^\Sum}			% set of sum views
\newcommand{\VM}{\V^\Max}			% set of max views
\newcommand{\VR}{\V^\Reg}			% set of regular views

\newcommand{\vc}{v^\Count}			% count view
\newcommand{\vs}{v^\Sum}			% sum view
\newcommand{\vm}{v^\Max}			% max view
\newcommand{\vr}{v^\Reg}			% regular view

%%%%%%%%%%%%%%%%%%%%%%%%%%%%%%%%%%%  Equivalence of Aggregate Queries

\newcommand{\wclass}[1]{[#1]_{\sl w}}		% equivalence class under weak equivalence

%%%%%%%%%%%%%%%%%%%%%%%%%%%%%%%%%%%  Abbreviations

\newcommand{\DW}{data warehouse}

%%%%%%%%%%%%%%%%%%%%%%%%
\newcommand{\Corr}[1]{#1}
\newcommand{\RM}[1]{}

%%%%%%%%%%%%%%%%%%%%%%%%%%%%%%%%%%%%%%%%%%%%%%%%%%%%%%%%%%%%%%%%%%%%%%%%
\newcommand{\sfq}{{\sf q}}
\newcommand{\ta}{{\sf ta}}
\newcommand{\salaries}{{\sf salaries}}

\newcommand{\Name}{{\sl Name}}
\newcommand{\CourseName}{{\sl Course\_Name}}
\newcommand{\JobType}{{\sl Job\_Type}}
\newcommand{\Amount}{{\sl Amount}}
\newcommand{\Sponsorship}{{\sl Sponsorship}}
\newcommand{\ISF}{\mbox{\tt 'Govt.'}}

%% file: title.tex
\title{Algorithms for Rewriting Aggregate Queries Using Views}
\author{{\bf Sara Cohen%\thanks{Contact author.}
}\\
	Computer Science Department\\
	The Hebrew University \\
	Jerusalem 91904, Israel \\
	sarina@cs.huji.ac.il  \\
%	Phone:  +972-2-658 677 \\
%	Fax:  +972-2-658 5439 \\
	\mbox{} \\
	{\bf Werner Nutt}\\
	German Research Center for Artificial Intelligence GmbH\\ 
	Stuhlsatzenhausweg 3 \\ 
	66123 Saarbr\"ucken, Germany \\
	Werner.Nutt@dfki.de \\
%	Phone: +49-681-302-5322 \\
%	Fax:   +49-681-302-5325 \\
	\mbox{} \\
	{\bf Alexander Serebrenik} \\
	Computer Science Department\\
	K. U. Leuven \\
	Celestijnenlaan 200A,   \\
	B-3001 Heverlee, Belgium \\
	Alexander.Serebrenik@cs.kuleuven.ac.be \\
%	Phone: +32-16-32-75-60 \\
%	Fax:   +32-16-32-79-96 
}
\date{}
\institute{}
\authorrunning{S.Cohen, W.Nutt, A.Serebrenik}
\maketitle
\begin{center}
Technical Report CW 292
\end{center}
\thispagestyle{empty}

%\begin{center}
%\fbox{
%\parbox{5in}{\em
%\begin{tabbing}
%{\bf Contact Author:} \\
%Sara Cohen \\
%Computer Science Department \\
%The Hebrew University, \\
%Jerusalem 91904, Israel \\ \\
%Email: sarina@cs.huji.ac.il \\
%Fax:  +972-2-658 5439 
%\end{tabbing}
%}}
%\end{center} 

%% file: abstract.tex
\begin{center}
{\LARGE Algorithms for Rewriting Aggregate Queries Using Views} \\
\mbox{} \\
%{\large Sara Cohen \ \ \ \ \ Werner Nutt \ \ \ \ \ Alexander Serebrenik}
\end{center}

%\vspace*{0.7cm}

% 163 words

\begin{abstract}
Queries involving aggregation are typical in database applications.
One of the main ideas to optimize the execution of an aggregate query
is to reuse results of previously answered queries.
This leads to the problem of rewriting aggregate queries
using views.
Due to a lack of theory, algorithms for this problem were
rather ad-hoc.  They were sound, but were not proven to be complete.

Recently we have given syntactic characterizations for the
equivalence of aggregate queries and applied them to decide when there
exist rewritings.  However, these decision procedures do not lend themselves 
immediately to an implementation.
In this paper, we present practical 
algorithms for rewriting queries with $\COUNT$ and $\SUM$.
Our algorithms are sound. They are also complete for important cases.
Our techniques can be used to improve well-known procedures
for rewriting non-aggregate queries. These procedures can then be
adapted to obtain algorithms for rewriting queries with $\MIN$ and $\MAX$.
The algorithms presented are a basis for realizing optimizers 
that rewrite queries using views. 

\end{abstract}

%% file: intro.tex
\section{Introduction}
%%%%%%%%%%%%%%%%%%%%%%%%%%%%%%%%%%%%%%%%%%%%%%%%%%%%%%%%%%%%%%%%%%%%%%%%%%%%%
Aggregate queries occur in many applications, such as data 
warehousing~\cite{Theodoratos:Sellis-D:W:Configuration-VLDB}, mobile computing~\cite{BI94}, 
and global information systems~\cite{LRO96}. The size of the database in these
applications is generally very large. Aggregation is often used in queries
against such sources as a means of reducing the granularity of data. 
The execution of aggregate queries tends to be time consuming and costly.
Computing one aggregate value often requires scanning many data items.
This makes query optimization a necessity.
A promising technique to speed up the execution of aggregate queries 
is to reuse the answers to previous queries to answer new queries. If the 
previous queries involved aggregation, the answers to them will tend to be 
much smaller than the size of the database. Thus, using their answers will
be much more efficient.

We call a reformulation of a query that uses other queries a 
{\em rewriting.}
Finding such rewritings is known %in the literature 
as the problem of 
{\em rewriting queries using views.}
In this phrasing of the problem, it is assumed that there is a set of
{\em views,} whose answers have been stored, or {\em materialized.}
Given a query, the problem is to find 
a rewriting, which is formulated 
in terms of the views and some database relations, such that 
evaluating the original query yields the same answers as evaluating
first the views and then the rewriting.

Rewriting queries using views has been studied for
non-aggregate queries
\cite{Levy:Et:Al-Reusing:Views-PODS}, 
and algorithms have been devised and implemented
\cite{Levy:Et:Al-Global:Information:Systems-JIIS,%
	Qian-Query:Folding-ICDE}.
For aggregate queries, the problem has been investigated mainly in
the special case of datacubes 
\cite{Harinarayan:Et:Al-Efficient:Data:Cubes-SIGMOD,%
	Dyreson-Incomplete:Datacube-VLDB}.
However, there is little theory for general aggregate queries, and the
rewriting algorithms that appear in the literature are by and large ad
hoc.  These algorithms are sound, that is, the reformulated queries
they produce are in fact rewritings, but there is neither a guarantee
that they output a rewriting whenever one exists, nor that they
generate all existing rewritings
\cite{Srivastava:Et:Al-Reusing:Views:With:Aggregates-VLDB,%
	Gupta:Et:Al-Generalized:Projections-VLDB}.

Recently, syntactic characterizations for the
equivalence of SQL queries with the aggregate operators $\MIN$,
$\MAX$, $\COUNT$, and $\SUM$ have been given
\cite{Nutt:Et:Al-Equivalences:Among:Aggregate:Queries-PODS}.
They have been applied to decide, given an aggregate query and a
set of views, whether there exists a rewriting, and 
whether a new query over views and base relations is a
rewriting \cite{Cohen:Et:Al-Rewriting:Aggregate:Queries-PODS}.

Using these characterizations, one can ``guess'' candidates
for rewritings and verify if they are in fact equivalent to the original 
query. However, this process is highly nondeterministic. Clearly, it is
more efficient to gradually build a candidate for rewriting in a way that
will ensure its being a rewriting. The characterizations do not immediately
yield practical algorithms of this sort. In fact, there are several subtle 
problems that must be dealt with in order to yield complete algorithms. 
%For example, we will show that in order to rewrite a query with comparisons it
%is necessary to assume that the comparisons are deductively closed.

In this paper, we show how to derive practical algorithms for rewriting
aggregate queries.
The algorithms are sound, i.e., they output rewritings.
We can also show that they are complete for important cases, which are
relevant in practice. In Section 2 we present a motivating example.
A formal framework for rewritings of aggregate queries is presented in Section 3.
In Section 4 we give algorithms for rewriting aggregate queries and in Section 5 we conclude. In Appendix~\ref{appendix} we demonstrate how queries written 
in SQL can be translated to our extended Datalog syntax and vice versa.

%% file: motivation.tex
\section{Motivation}
	\label{sec-motivation}
%%%%%%%%%%%%%%%%%%%%%%%%%%%%%%%%%%%%%%%%%%%%%%%%%%%%%%%%%%%%%%%%%%%%%%%%%%%%%

We discuss an example that illustrates the rewriting problem 
for aggregate queries. All the examples in this paper are written 
using an extended Datalog syntax. This syntax is more abstract and 
concise than SQL. In Section~\ref{sec-framework} we present a
formal definition of the Datalog syntax. In Appendix~\ref{appendix} we 
describe how queries written in SQL can be translated to our Datalog syntax
and vice versa.

The following example models exactly the payment
policy for teaching assistants at the Hebrew University in Jerusalem.
There are two tables with relations pertaining to salaries of teaching
assistants:
\begin{center}
{\tt
	ta(name,course\_name,job\_type) {\rm and } 
	salaries(job\_type,sponsorship,amount)}.
	%}
\end{center}
At the Hebrew University, there may be many teaching assistants in a
course.
Each TA has at least one {\tt job\_type} in the course he assists.
For example, he may give lectures or grade exercises.
Teaching assistants are financed by different sources, like science
foundations and the university itself.
For each job type, each sponsor gives a fixed amount.
Thus, a lab instructor may receive \$600 per month from the university
and \$400 from a government science foundation.

We suppose that there are two materialized views.
In the first one of them, {\tt v\_positions\_per\_type}, we compute the number
of positions of each type held in the university. 
In the second view, {\tt v\_salary\_for\_ta\_job} we compute the total salary 
for each type of position.  We express aggregate queries with an
extended Datalog notation, where
in the head we separate grouping variables and aggregate terms by a
semicolon:
\begin{eqnarray*}
{\tt v\_positions\_per\_type}(j;\COUNT) & \qif & {\tt ta}(n,c,j) \\
{\tt v\_salary\_for\_ta\_job}(j;\SUM(a)) & \qif & {\tt salaries}(j,s,a).
\end{eqnarray*}

\noindent
 In Subsection \ref{ssec-aggr:queries} we define a semantics for such
Datalog queries that identifies them with SQL queries where the
attributes in the {\tt GROUP BY} clause and those in the {\tt SELECT} clause
coincide.  
The grouping variables correspond to those attributes.

In the following query we calculate the total amount of money spent on each
job position:
\begin{center}
$  q(j;\SUM(a)) \qif {\tt ta}(n, c, j) \AND {\tt salaries}(j, s, a) $
\end{center}
An intelligent query optimizer could now reason that for each type of job
we can calculate the total amount of money spent on it 
if we multiply the salary that one TA receives for such a job
by the number of positions of that type.
The two materialized views contain information that can be combined to yield an answer to our 
query. The optimizer can formulate a new query that only
accesses the views and does not touch the tables in the database:
\begin{center}
$r(j'; a' * {\it cnt}) \qif {\tt v\_positions\_per\_type}(j';{\it cnt}) \AND 
{\tt v\_salary\_for\_ta\_job}(j';a') $
\end{center}
In order to evaluate the new query, we no longer need to look up all the
teaching assistants nor all the financing sources.
Thus, probably, the new query can be executed more efficiently.

In this example, we used our common sense in two occasions.
First, we gave an argument why evaluating the original query yields
the same result as evaluating the new query that uses the views.
Second, because we understood the semantics of the original query and
the views, we were able to come up with a reformulation of the query
over the views.
Thus, we will only be able to build an optimizer that can rewrite
aggregate queries,
if we can provide answers to the following two questions.
\begin{itemize}
\item
{\bf Rewriting Verification:}
How can we prove that a new query, which uses views, produces the same
results as the original query?
\item
{\bf Rewriting Computation:}
How can we devise an algorithm that systematically and efficiently
finds all rewritings?
\end{itemize}

If efficiency and completeness cannot be achieved at the same time, we
may have to find a good trade-off between the two requirements.

%% file: framework.tex
\section{A Formal Framework}
	\label{sec-framework}
%%%%%%%%%%%%%%%%%%%%%%%%%%%%%%%%%%%%%%%%%%%%%%%%%%%%%%%%%%%%%%%%%%%%%%%%

In this section we define the formal framework in which we study 
rewritings of aggregate queries.
We extend the well-known Datalog syntax for non-aggregate
queries~\cite{Ullman-Database:And:K:B:Systems:II}
so that it covers also aggregates.
These queries express nonnested SQL queries
without a {\tt HAVING} clause and with the aggregate operators $\MIN$, $\MAX$,
$\COUNT$, and $\SUM$.  
A generalization to queries with the constructor {\tt UNION} is possible,
but beyond the scope of this paper.
For queries with arbitrary nesting and negation no rewriting
algorithms are feasible, since equivalence of such queries
is undecidable.

\subsection{Non-aggregate Queries}
	\label{ssec-non:agg}
%%%%%%%%%%%%%%%%%%%%%%%%%%%%%%%%%%%%%%%%%%%%%%%%%%%%%%%%%%%%%%%%%%%%%%%%%%%%%%%%%%%%%%%

We recall the Datalog notation for conjunctive queries
and extend it to aggregate queries.

A~{\em term\/} (denoted  as $s$, $t$)
is either a variable (denoted  as $x$, $y$, $z$)
or a constant.
A {\em comparison\/} has the form $s_1 \osps{\rho} s_2$, 
where $\rho$ is either $<$ or $\leq$. %, $>$, and $\geq$.%
% \footnote{We use the notation $s = t$ as abbreviation 
%          for the conjunction $s \leq t \AND t \leq s$.}
If $C$ and $C'$ are conjunctions of comparisons, we
write $C \models C'$ if $C'$ is a consequence of $C$.
We assume all comparisons range over the rationals.

We denote predicates as $p$, $q$, $r$.
A {\em relational atom\/} has the form $p(\dd s k)$. 
Sometimes we write $p(\tpl s)$,
where $\tpl s$ denotes the tuple of terms $\dd s k$.
An {\em atom\/} (denoted as $a$,~$b$) 
is either a relational atom or a comparison.

A {\em conjunctive query\/} is an expression of the form 
$q(\dd x k) \qif  a_1\AND\cdots\AND a_n$.
The atom $q(\dd x k)$ is called the {\em head\/} of the query.
The atoms $a_1,\ldots,a_n$ form the query {\em body.} 
They can be relational or comparisons.
If the body contains no comparisons, then the query is {\em relational}.
A query is {\em linear\/} if it does not contain two relational atoms with the 
same predicate symbol.
We abbreviate a query as $q(\tpl x) \qif  B(\tpl s)$,
where $B(\tpl s)$ stands for the body and $\tpl s$ for
the terms occurring in the body.
Similarly, we may write a conjunctive query as
$q(\tpl x) \qif R(\tpl s) \AND C(\tpl t)$,
in case we want to distinguish between the relational atoms and the
comparisons in the body,
or, shortly, as $q(\tpl x) \qif R \AND C$.
The variables appearing in the head are called 
{\em distinguished variables},
while those appearing only in the body are called 
{\em nondistinguished variables.} Atoms containing at least one nondistinguished
variable are called {\em nondistinguished atoms.}
By abuse of notation, we will often refer to a query by its head 
$q(\tpl x)$ or simply by the predicate of its head $q$.

A database $\D$ contains for every predicate symbol $p$ a relation $\DBD p$,
that is, a set of tuples.
Under {\em set semantics,} a conjunctive query $q$ defines a new
relation $\DBD q$, which consists of all the answers that $q$ produces
over $\D$.
Under {\em bag-set semantics,} $q$ defines a multiset or {\em bag\/}
$\DBDBAG q$ of tuples. % for each database $\D$.
The bag $\DBDBAG q$ contains the same tuples as the relation $\DBD q$,
but each tuple occurs as many times as it can be derived over $\D$
with $q$ \cite{Chaudhuri:Vardi-Real:Conjunctive:Queries-PODS}.

Under set-semantics, 
two queries $q$ and $q'$ are {\em equivalent\/} 
if for every database, they return the same set as a result.
Analogously, we define equivalence under bag-set-semantics.

\subsection{Aggregate Queries}
	\label{ssec-aggr:queries}
%%%%%%%%%%%%%%%%%%%%%%%%%%%%%%%%%%%%%%%%%%%%%%%%%%%%%%%%%%%%%%%%%%%%%%%%%%%

We now extend the Datalog syntax so as to capture also queries with
{\tt GROUP BY} and aggregation.
We assume that queries have only one aggregate term.
The general case can easily be reduced to this one 
\cite{Cohen:Et:Al-Rewriting:Aggregate:Queries-PODS}.
We are interested in queries with the aggregation functions $\COUNT$,
$\SUM$, $\MIN$ and $\MAX$.
Since results for $\MIN$ are analogous to those for $\MAX$, 
we do not consider $\MIN$.
Our function $\COUNT$ is analogous to the function {\tt COUNT(*)} of
SQL.

An {\em aggregate term\/} is an expression built up using variables,
the operations addition and multiplication, 
and aggregation functions.%
\footnote{%
	This definition blurs the distinction between the function as
	a mathematical object and the symbol denoting the function.
	However, a notation that takes this difference into account
	would be cumbersome.}
For example, $\COUNT$ and $\SUM(z_1 * z_2)$, are aggregate terms.
We use $\kappa$ as abstract notations for aggregate
terms.
If we want to refer to the variables occurring in an aggregate term, 
we write $\kappa(\tpl y)$,
where $\tpl y$ is a tuple of distinct variables. Note that $\tpl y$ 
is empty if $\kappa$ is the $\COUNT$ aggregation function. 
Terms of the form $\COUNT$, $\SUM(y)$ and $\MAX(y)$ 
are {\em elementary aggregate terms}. 
Abstractly, elementary aggregate terms are denoted as $\alpha(y)$,
where $\alpha$ is an aggregation function.

An aggregate term $\kappa(\tpl y)$ naturally gives rise to a function
$f_{\kappa(\tpl y)}$ that maps multisets of tuples of numbers to numbers.
For instance, $\SUM(z_1 * z_2)$ describes the aggregation 
function $f_{\SUM(z_1 *
z_2)}$ that maps any multiset $M$ of pairs of numbers $(m_1,m_2)$ to
$\sum_{(m_1,m_2)\in M} m_1*m_2$. 

An {\em aggregate query\/} is a conjunctive query augmented by an
aggregate term in its head. Thus, it has the form
$q(\tpl x;\kappa(\tpl y))\qif B(\tpl s)$.
We call $\tpl x$ the {\em grouping variables\/} of the query. 
Queries with elementary aggregate terms are {\em elementary queries}.
If the aggregation term in the head of a query has the form $\alpha(y)$,
we call the query an {\em $\alpha$-query\/}
(e.g., a $\MAX$-query).
In this paper we are interested in rewriting elementary queries using
elementary views. 
However, as the example in Section~\ref{sec-motivation} shows, even
under this restriction the rewritings may not be elementary.

We now give a formal definition of the semantics of aggregate queries. 
Consider the query $q(\tpl x;\kappa(\tpl y))\qif B(\tpl s)$.
For a database $\D$, the query yields a new relation $\DBD q$.
To define the relation $\DBD q$, we proceed in two steps.
We associate to $q$ a non-aggregate query, $\core q$, 
called the {\em core\/} of $q$, which is defined as
$\core q(\tpl x,\tpl y) \qif B(\tpl s)$.
The core is the query that returns all the values that are amalgamated
in the aggregate.
Recall that under bag-set-semantics, the core returns over $\D$ a bag
$\DBDBAG{\core q}$ of tuples $(\tpl d,\tpl e)$.
For a tuple of constants $\tpl d$ of the same length as $\tpl x$, let
\begin{center}$
	\Gamma_{\tpl d}  := 
	\bigbag{ \tpl e \bigmid (\tpl d, \tpl e) \in \DBDBAG{\core q}}$.
\end{center}
That is, the bag $\Gamma_{\tpl d}$ is obtained by first grouping
together those answers to $\core q$ that return $\tpl d$ for the 
grouping terms,
and then stripping off from those answers the prefix $\tpl d$.
In other words, $\Gamma_{\tpl d}$ is the multiset of $\tpl y$-values that
$\core q$ returns for $\tpl d$.
The result of evaluating $q$ over $\D$ is
\begin{center}
$        \DBD q  :=  
        \{ (\tpl d, e) \mid 
        \Gamma_{\tpl d} \neq \eset {\rm \ and\ }
                 e = f_{\kappa(\tpl y)}(\Gamma_{\tpl d}) \} $.
\end{center}
Intuitively, whenever there is a nonempty group of answers
with index $\tpl d$, 
then we apply the aggregation function $f_{\kappa(\tpl y)}$
to the multiset of $\tpl y$-values of that group.

Again, two aggregate queries $q$ and $q'$ are {\em equivalent\/} if 
$\DBD q = \DBD {q'}$ for all databases $\D$.

\subsection{Equivalence Modulo a Set of Views}
        \label{ssec-equiv:modulo:views}
%%%%%%%%%%%%%%%%%%%%%%%%%%%%%%%%%%%%%%%%%%%%%%%%%%%%%%%%%%%%%%%%%%%%%%%
Up until now, we have defined equivalence of aggregate queries and
equivalence of non-aggregate queries under set and
bag-set-semantics.
However, the relationship between a query $q$ and a rewriting $r$ of $q$
is not equivalence of queries,
because the view predicates occurring in $r$ are not regular database 
relations,
but are determined by the base relations indirectly.
In order to take this relationship into account, we define %give a definition of 
equivalence of queries modulo a set of views.

We consider aggregate queries that use predicates 
both from $\R$, a set of base relations,  
and~$\V$, a set of view definitions.
For a database $\D$, let $\EXTDB\D\V$ be the database that extends $\D$
by interpreting every view predicate $v\in\V$ as the relation~$\DBD v$.
If $q$ is a query that contains also predicates from~$\V$,
then $\DB q {\EXTDB \D\V}$ is the relation that results from evaluating
$q$ over the extended database $\EXTDB \D\V$.
If $q$, $q'$ are two aggregate queries using predicates from $\R\cup\V$,
we define that $q$ and $q'$ are {\em equivalent modulo $\V$},
written $q \equivv q'$,
if $\DB q {\EXTDB \D\V} = \DB{q'}{\EXTDB \D\V}$ for all databases $\D$.

\subsection{General Definition of Rewriting}
%%%%%%%%%%%%%%%%%%%%%%%%%%%%%%%%%%%%%%%%%%%%%%%%%%%%%%%%%%%%%%%%%%%%%%%%%%%%%%
We give a general definition of rewritings.
Let $q$ be a query, 
$\V$ be a set of views over the set of relations $\R$, and 
$r$ be a query over $\V \cup \R$.
All of $q$, $r$, and the views in $\V$ may be aggregate queries or not.
Then we say that 
   $r$ is a {\em rewriting\/} of $q$ using $\V$ if 
   $q \equiv_{\V} r$ and $r$ contains only atoms with predicates from $\V$.
If $q \equiv_{\V} r$ and $r$ contains at least one atom with a predicate from $\V$
we say that $r$ is a {\em partial rewriting of $q$ using $\V$\/}.

Now we can reformulate the intuitive questions we asked in the end of the 
Section~\ref{sec-motivation}.
\begin{itemize}
\item
{\bf Rewriting Verification:}
Given queries $q$ and $r$, and a set of views $\V$, 
check whether $q\equivv r$. 
\item
{\bf Rewriting Computation:}
Given a query $q$ and a set of views $\V$,  
find all (some) rewritings or partial rewritings of $q$. 
\end{itemize}

%% file: rewritings.tex
\section{Rewritings of Aggregate Queries}
%%%%%%%%%%%%%%%%%%%%%%%%%%%%%%%%%%%%%%%%%%%%%%%%%%%%%%%%%%%%%%%%%%%%%%%%%%%%%%%
We now present techniques for rewriting aggregate queries.
Our approach will be to generalize the known techniques for
conjunctive queries.
Therefore, we first give a short review of the conjunctive case
and then discuss in how far aggregates give rise to more
complications.

\subsection{Reminder: Rewritings of Relational Conjunctive Queries}
\label{conj:rew}
%%%%%%%%%%%%%%%%%%%%%%%%%%%%%%%%%%%%%%%%%%%%%%%%%%%%%%%%%%%%%%%%%%%%%%%%%%%%%%%%
We review the questions related to rewriting relational conjunctive
queries.
Suppose, we are given a set of conjunctive queries $\V$, the views, 
and another conjunctive query $q$.
We want to know whether there is a rewriting of $q$ 
using the views in $\V$.

The {\em first question\/} that arises is, what is the {\em
language\/} for expressing rewritings?
Do we consider arbitrary first order formulas over the view predicates
as candidates, or recursive queries, or do we restrict ourselves
to conjunctive queries over the views?
Since reasoning about queries in the first two languages is
undecidable, researchers have only considered conjunctive rewritings.% 
\footnote{It is an interesting theoretical question, which as yet has
	  not been resolved, whether more expressive languages give
	  more possibilities for rewritings.  It is easy to show, at
	  least, that in the case at hand allowing also disjunctions
	  of conjunctive queries as candidates does not give more
	  possibilities than allowing only conjunctive queries.}
Thus, a candidate for  rewriting  $q(\tpl x)$ has the form
$r(\tpl x) \qif v_1(\theta_1 \tpl x_1) \AND \ldots \AND 
                   v_n(\theta_n \tpl x_n)$,
where the $\theta_i$'s are substitutions that instantiate the view
predicates $v_i(\tpl x_i)$.

The {\em second question\/} is whether we can {\em reduce reasoning\/}
about the query $r$, which contains view predicates, to reasoning
about a query that has only base predicates.
To this end, we {\em unfold\/} $r$.
That is, we replace each view atom $v_i(\theta_i\tpl x_i)$, 
with the instantiation $\theta_i B_i$ of the body of $v_i$,
where $v_i$ is defined as $v_i(\tpl x_i)\qif B_i$.
We assume that the nondistinguished variables in different occurrences
of the bodies are distinct.
We thus obtain the unfolding $\unf r$ of $r$, for which the Unfolding Theorem holds, $\unf r(\tpl x)\qif \theta_1 B_1\AND\ldots\AND\theta_n B_n$.

\begin{theorem}[Unfolding Theorem]
Let $\V$ be a set of views, $r$ a query over $\V$, and 
$\unf r$ be the unfolding of $r$.
Then $r$ and $\unf r$ are equivalent modulo $\V$, that is,
\begin{center}
		$r\equivv \unf r$.
\end{center}
\end{theorem}

The {\em third question\/} is how to check whether $r$ is a rewriting
of $q$, that is, whether $r$ and $q$ are 
{\em equivalent modulo $\V$\/}.
This can be achieved by checking
whether $\unf r$ and $q$ are set-equivalent:
if $\unf r\equiv q$, then the Unfolding Theorem implies
$r \equivv q$.
Set-equivalence of conjunctive queries can be decided syntactically by
checking whether there are \hom s in both directions
\cite{Ullman-Database:And:K:B:Systems:II}.

%% file: count_views.tex
\newcommand{\ndv}{{\sl ndv}}
\subsection{Rewritings of Count-queries}
\label{count:rewriting}

When rewriting $\COUNT$-queries, we must deal with the same questions that
arose when rewriting conjunctive queries. Thus, we first define the language
for expressing rewritings. Even if we restrict the language to conjunctive 
aggregate queries over the views, we still must decide on two additional 
issues. First, which types of aggregate views are useful for a rewriting?
Second, what will be the aggregation term in the head of the rewriting?  
A $\COUNT$-query is sensitive to multiplicities, and $\COUNT$-views are the
only type of aggregate views that do not 
lose multiplicities.%
\footnote{Although $\SUM$-views are sensitive
to multiplicities (i.e., are calculated under bag-set-semantics), they lose
these values. For example, $\SUM$-views ignore occurrences of zero values.}
Thus, the natural answer to the first question is to use only $\COUNT$-views
when rewriting $\COUNT$-queries. 
We show in the following example that there are an infinite number of aggregate
terms that can be usable in rewriting a $\COUNT$-query.

\begin{example}
\label{example:power:rewriting}
Consider the query
\begin{eqnarray*}
{\tt q\_positions\_per\_type}(j;\COUNT) \qif {\tt ta}(n,c,j)
\end{eqnarray*}

in which we compute the number of positions of each type held in the 
university. Recall the view {\tt v\_positions\_per\_type} defined in 
Section~\ref{sec-motivation}.
It is easy to see that both of the following are rewritings of 
{\tt q\_positions\_per\_type}:
\begin{eqnarray*}
r_1(j';z) & \qif &{\tt v\_positions\_per\_type}(j';z) \\
r_2(j';\sqrt{z_1 * z_2}) 
	& \qif & {\tt v\_positions\_per\_type}(j';z_1) \AND
		   {\tt v\_positions\_per\_type}(j';z_2). 
\end{eqnarray*}

By adding additional view atoms and adjusting the power of the root
we can create infinitely many different rewritings
of {\tt q\_positions\_per\_type}. It is natural to create only $r_1$ as
 a rewriting of $q$. In fact, only for $r_1$ will the Unfolding Theorem hold. 
%However, when creating 
%rewritings automatically we must restrict the
%aggregate term in the head of the rewriting in order to prevent deriving 
%infinitely many rewritings. 
\end{example}

We define a {\em candidate\/} for a rewriting of $q(\tpl x;\COUNT)\qif R\AND C$ as 
a query having the form
\begin{center}
$ r(\tpl x; \SUM({\displaystyle\prod_{i=1}^n} z_i))
        \qif    \vc_1(\theta_1 \tpl x_1; z_1) \AND \ldots \AND 
                \vc_n(\theta_n \tpl x_n; z_n) \AND C'$,
\end{center}
where $\vc_i$ are $\COUNT$-views, possibly with comparisons, defined as 
$\vc_i(\tpl x_i; \COUNT)\qif B_i$ and $z_i$ are variables not appearing elsewhere
in the body of $r$. We call $r$ a $\COUNT$-rewriting candidate. 

Note that it is possible to omit the summation if the values of $z_i$
are functionally dependent on the value of the grouping variables
$\tpl x$.
This is the case, if only grouping variables appear as $\theta_i x_i$
in the heads of the instantiated views.
Then the summation is always over a singleton group.

After presenting our rewriting candidates we now show how we can reduce 
reasoning about rewriting candidates, to reasoning about conjunctive 
aggregate queries. We use a similar technique to that shown in 
Subsection~\ref{conj:rew}. 
In the unfolding, we replace the view atoms of the rewriting with the 
appropriate instantiations of their bodies, and we replace the 
aggregate term in the rewriting with the term $\COUNT$.  
Thus, we obtain as the unfolding $\unf r$ of $r$ the query
\begin{center}
$\unf r(\tpl x;\COUNT) \qif 
	\theta_1 B_1 \AND \ldots \AND \theta_n B_n \AND C'$.
\end{center}
 
In~\cite{Cohen:Et:Al-Rewriting:Aggregate:Queries-PODS},
it has been proven that for $\unf r$ the Unfolding Theorem holds, 
i.e., $r\equivv \unf r$.
Moreover, it has been shown that this definition of unfolding uniquely
determines the aggregation function in the head of our candidates.
That is, summation over products of counts is the only aggregation
function for which the Unfolding Theorem holds if $\unf r$ is defined
as above.
Now, in order to verify that $r$ is a rewriting of $q$, we can
check that $\unf r$ is equivalent to $r$, 
without taking into account the views any more.

We now present an algorithm that finds a rewriting for a 
$\COUNT$-query using views.
Our approach can be thought of as reverse engineering. 
We have characterized the ``product'' that we 
must create, i.e., a rewriting, and we now present an 
automatic technique for producing it. 

In~\cite{Nutt:Et:Al-Equivalences:Among:Aggregate:Queries-PODS}, 
a sound and complete characterization of equivalence of 
conjunctive $\COUNT$-queries with comparisons has been given.
The only known algorithm that checks equivalence of conjunctive 
$\COUNT$-queries creates an exponential blowup of the queries. 
Thus, it is difficult to present a tractable algorithm for 
computing rewritings.
However, it has been
shown~\cite{Chaudhuri:Vardi-Real:Conjunctive:Queries-PODS,%
Nutt:Et:Al-Equivalences:Among:Aggregate:Queries-PODS} that two
relational $\COUNT$-queries are equivalent if and only if they are isomorphic.
In addition, equivalence of linear $\COUNT$-queries with comparisons 
is isomorphism of the 
queries~\cite{Nutt:Et:Al-Equivalences:Among:Aggregate:Queries-PODS}. 
Thus, we will give a sound, complete, and tractable algorithm
for computing rewritings of relational $\COUNT$-queries and of
linear $\COUNT$-queries. This algorithm is sound
and tractable for the general case, but is not complete. 

We discuss when a view 
$v(\tpl u;\COUNT)\qif R_v \AND C_v$, 
instantiated by $\theta$, 
is usable in order to rewrite a query $q(\tpl x;\COUNT)\qif R \AND C$,
that is, when the instantiated view can occur in a partial rewriting. 
By the characterization of equivalence for relational and linear queries,
a rewriting of $q$ is a query $r$ that when unfolded yields a 
query isomorphic to $q$.
Thus, in order for $\theta v$, to be usable,  
$\theta R_v$ must ``cover'' some part of $R$. 
Therefore, $\theta v$ is usable for rewriting $q$  only if there exists an isomorphism,
$\varphi$, from $\theta R_v$ to $R'\subseteq R$.  Note that we can assume, w.l.o.g.\ 
that $\varphi$ is the identity mapping on the distinguished variables of $v$. 
We would like to replace $R'$ with 
$\theta v$ in the body of $q$ in order to derive a partial rewriting of $q$. 
This cannot
always be done. Observe that after replacing $R'$ with $\theta v$, variables 
that appeared in $R'$ and do not appear in $\theta \tpl u$ (i.e., the nondistinguished variables
in $v$) are not accessible anymore. 
Thus, we can only perform the replacement if these variables do not appear anywhere 
else in $q$, in $q$'s head or body.  
We capture this property by defining that 
$v(\tpl u;\COUNT) \qif R_v \AND C_v$ is 
{\em $R$-usable under $\theta$ w.r.t.\ $\varphi$\/} if 
\begin{enumerate}
\item $\varphi\theta R_v$ is isomorphic to a subset $R'$ of $R$, and
\item every variable that occurs both in $R'$ and in $R\setminus R'$
    must occur in $\theta\tpl u$.
\end{enumerate}
We denote this fact as {\sf $R$-usable($v$, $\theta$, $\varphi$)}.
Clearly, there is a partial rewriting using $v$ 
of $q(\tpl x;\COUNT)\qif R \AND C$ 
 only if 
$v(\tpl u;\COUNT) \qif R_v \AND C_v$ is 
$R$-usable under $\theta$ w.r.t.\ some~$\varphi$.

\begin{example}
Consider the following query that computes the number of sponsors for 
each assistant in the database course 
\begin{eqnarray*}
 {\tt q\_db\_ta\_sponsors}(n;\COUNT) \qif {\tt ta}(n,{\tt Database},j) 
	\AND {\tt salaries}(j,s,a). 
\end{eqnarray*}

We suppose that we have a materialized view that computes the number of jobs 
that each teaching assistant has in each course that he assists
\begin{eqnarray*}
 {\tt v\_jobs\_per\_ta}(n',c';\COUNT) \qif {\tt ta}(n',c',j').
\end{eqnarray*}

In order to use {\tt v\_jobs\_per\_ta} in rewriting {\tt q\_db\_ta\_sponsors}
 we must find an instantiation $\theta$ such that
$\theta {\tt ta}(n',c',j')$ covers some part of the body of 
{\tt q\_db\_ta\_sponsors}. 
Clearly, $\theta {\tt ta}(n',c',j')$ can cover only 
${\tt ta}(n,{\tt Database},j)$. We take, $\theta=\{n'/n,c'/{\tt Database}\}$ 
and thus, $\varphi=\{n/n,j'/j \}$. However, $j$ appears in 
${\tt ta}(n,{\tt Database},j)$ and not in the head of 
$\theta {\tt v\_jobs\_per\_ta}$ and therefore, $j$ is not accessible 
after replacement. Note that $j$  appears in {\tt salaries} and
thus, {\tt v\_jobs\_per\_ta} is not $R$-usable in rewriting 
{\tt q\_db\_ta\_sponsors}. 
\end{example}

For our algorithm to be complete for linear queries, 
the set of comparisons in the query to be rewritten has to be
deductively closed (see Example~\ref{deductive:example}). 
The deductive closure of a set of comparisons can be computed in
polynomial time~\cite{Klug-Conjunctive:Queries:Inequalities-JACM}.
In addition, it must hold that $C\models \varphi(\theta C_v)$, thus, 
the comparisons inherited from $v$ are weaker than those in $q$.
For a rewriting using $\theta v$ to exist it must be possible to
strengthen $\varphi(\theta C_v)$ by additional comparisons $C'$ 
so that $\varphi(\theta C_v) \AND C'$ is equivalent to $C$.
We have seen that when replacing $R'$ with $\theta v$ we lose access
to the nondistinguished variables in $v$.
Therefore, it is necessary for the comparisons in 
$\varphi(\theta C_v)$ 
to imply all the comparisons in $q$ that contain an image of a
nondistinguished variable in $v$.
Formally, let $\ndv(v)$ be the set of nondistinguished variables 
in $v$.
Let $C^{\varphi(\theta \ndv(v))}$ consist of those comparisons in 
$C$ that contain variables in $\varphi(\theta \ndv(v))$.
Then, in order for $\theta v$ to be usable in a partial rewriting,
$C_v\models C^{\varphi(\theta \ndv(v))}$ must hold. 
If this condition and $C\models \varphi(\theta C_v)$ hold, then we say
that {\em $v$ is $C$-usable under
$\theta$ w.r.t.\ $\varphi$} and write
{\sf $C$-usable($v$, $\theta$, $\varphi$)}.

We summarize the discussion in a theorem.
\begin{theorem}
Let $q(\tpl x;\COUNT)\qif R\AND C$ be a $\COUNT$-query 
whose set of comparisons $C$ is deductively closed, 
and let 
$v(\tpl u;\COUNT) \qif R_v\AND C_v$ be a $\COUNT$-view.
There exists a partial rewriting of $q$ using $v$
if and only if 
there is a $\varphi$ such that 
{\sf $R$-usable($v$, $\theta$, $\varphi$)} and 
{\sf $C$-usable($v$, $\theta$, $\varphi$)}.
\end{theorem}

% One can show that a partial rewriting 
% of $q(\tpl x;\COUNT)\qif R\AND C$ 
% using $v(\tpl u;\COUNT) \qif R_v\AND C_v$ exists
% if and only if 
% there is a $\varphi$ such that 
% {\sf $R$-usable($v$, $\theta$, $\varphi$)} and 
% {\sf $C$-usable($v$, $\theta$, $\varphi$)}.

\begin{example}
The following query computes for each job the number of mediocre sponsors, 
i.e., the number of sponsors who give an amount that is greater than \$200 and less than \$600.
\begin{eqnarray*}
 {\tt q\_mediocre\_sponsor}(j;\COUNT) \qif {\tt salaries}(j,s,a) \AND a>200 \AND a<600. 
\end{eqnarray*}

The view
\begin{eqnarray*}
{\tt v\_all\_sponsor}(j';\COUNT) \qif {\tt salaries}(j',s',a') \AND a'>0 
\end{eqnarray*}

\noindent computes the number of sponsors for each job.
In order to use {\tt v\_all\_sponsor} in rewriting {\tt q\_mediocre\_sponsor}
we clearly must take $\theta = \{j'/j\}$ and $\varphi = \{j/j,s'/s,a'/a\}$. 
It holds 
that $\{a>200 \AND a<600\} \models \{\varphi\theta (a'>0)\}$.
Observe that $a'$ is a nondistinguished 
variable in {\tt v\_all\_sponsor} and $a'$ is mapped to $a$ by 
$\varphi\theta$. Thus, in order for {\tt v\_all\_sponsor} to be  $C$-usable for rewriting
{\tt q\_mediocre\_sponsor} it must hold that 
 $\{\varphi\theta (a'>0) \} \models \{a>200 \AND a<600\}$. This does not hold. Therefore,  
{\tt v\_all\_sponsor} is not $C$-usable for rewriting
{\tt q\_mediocre\_sponsor}.
\end{example}

\newcommand{\nc}{{\sl Not\_Covered}}
\newcommand{\rew}{{\sl Rewriting}}

We present an algorithm for computing rewritings of conjunctive $\COUNT$-queries in 
Figure~\ref{fig:count:query:rew-algo}.
The underlying idea is to incrementally cover the body of the query 
by views until no atom is left to be covered.
The algorithm nondeterministically chooses a view $v$ and an instantiation $\theta$, such 
that $v$ is both $R$-usable and $C$-usable under $\theta$. 
If the choice fails, backtracking  is performed.

When the while-loop is completed, the algorithm returns a rewriting. 
By backtracking we can find additional rewritings. 
Of course, the nondeterminism in choosing the views can be further
reduced, for instance, by imposing an ordering on the atoms in the
body of the query and by trying to cover the atoms according to that
ordering.
Note, that the same algorithm may be used to produce partial
rewritings if we relax the termination condition of the while-loop.
This will similarly hold for subsequent algorithms presented.

\begin{figure*}[htb]
\def\baselinestretch{1.2}\small\normalsize
\begin{center}
\fbox{
\parbox{6in}{%%%\tt
\begin{tabbing}
AAAA \= AAAAA\= AAAAA\= AAAAA\= AAAAA\= AAAAA\= AAAAA\=\kill
{\bf Algorithm} \>\> {\sf Count\_Rewriting}\\
{\bf Input}     \>\> A query $q(\tpl x;\COUNT)\qif R\AND C$ and a set of views $\V$\\
{\bf Output}    \>\> A rewriting $r$ of $q$. \\
\\
\ \,(1) \> $\nc:=R$.\\
\ \,(2) \> $\rew:=\emptyset$.\\
\ \,(3) \> $n:=0$.\\
\ \,(4) \> {\bf While} $\nc\neq\emptyset$ {\bf do}: \\
\ \,(5) \>	\> {\bf Choose} a view $v(\tpl x';\COUNT)\qif R'\AND C'$ in $\V$. \\
\ \,(6) \>	\> {\bf Choose} an instantiation, $\theta$, and an isomorphism $\varphi$, \\
    \>	\>	\>such that {\sf $R$-usable($v$, $\theta$, $\varphi$)} and 
				{\sf $C$-usable($v$, $\theta$, $\varphi$)}. \\
\ \,(7) \>	\> {\bf For each} atom $a\in R'$ {\bf do}: \\
\ \,(8) \>	\> 	\> {\bf If} $a$ is a nondistinguished atom, {\bf then} \\
\ \,(9) \>	\>	\> 	\> {\bf Remove} $\varphi(\theta a)$ {\bf from} $R$.\\
(10) \>	\>	\>	\> {\bf If} $\varphi(\theta a)\not\in\nc$ {\bf then fail}.\\
(11) \>	\>	\> {\bf Remove} $\varphi(\theta a)$ {\bf from} $\nc$.\\
(12) \> 	\> {\bf Remove} {\bf from} $C$ comparisons containing a variable in $\varphi(\theta R')$, \\
     \> 	\> \> but not in $\theta \tpl x'$\\
(13) \>	\> {\bf Increment} $n$. \\
(14) \>	\> {\bf Add} $v(\theta\tpl x'; z_n))$ {\bf to} $\rew$, where $z_n$ is a fresh variable. \\
(15) \> {\bf Return} $r(\tpl x;\SUM(\prod_{i=1}^{n} z_i))\qif \rew \AND C$.
\end{tabbing}
}}
\caption{Count Query Rewriting Algorithm}
\label{fig:count:query:rew-algo}
\end{center}
\def\baselinestretch{2.0}\small\normalsize
\end{figure*}

We note the following. In Line 9, $R$ is changed and thus, $q$ is also
changed. 
Therefore, at the next iteration of the while-loop 
we check whether $v$ is $R$-usable under $\theta$ to 
rewrite the updated version of $q$ (Line 6). 
Thus, in each iteration of the loop, 
additional atoms are covered. 
In Line 10, the algorithm checks if a nondistinguished atom is
already covered. 
If so, then the algorithm must fail, i.e., backtrack, as explained 
above.

Observe that we modify $C$ in Line 12. 
We remove from $C$ its comparisons  containing a variable 
that is not accessible after replacing the appropriate subset of $R$ 
by the appropriate instantiation of $v$. 
These comparisons are not lost because $v$ is $C$-usable.
The comparisons remaining in $C$ are needed to strengthen those
inherited from the views such that they are equivalent to the
comparisons in the query to be rewritten.

{\sf Count\_Rewriting} is both sound and complete
for linear queries and relational queries and is sound, 
but not complete, for arbitrary queries. 
Our algorithm runs in nondeterministic polynomial time by guessing views 
and instantiations 
and verifying in polynomial time that the obtained result is a rewriting.
For relational queries this is optimal, since checking whether 
there exists a $\theta$ such
that $v$ is $R$-usable under $\theta$ is $\NP$-hard, which can be
shown by a reduction of the graph matching problem.
Since for linear queries $q$ and views $v$
the existence of $\theta$ and $\varphi$ such that 
{\sf $R$-usable($v$, $\theta$, $\varphi$)} and 
{\sf $C$-usable($v$, $\theta$, $\varphi$)} can be decided in
polynomial time,
one can obtain a polynomial time variant of the algorithm that computes
partial rewritings in the linear case.

\begin{theorem}{\bf (Soundness and Completeness of Count Rewriting)}
Let $q$ be a $\COUNT$-query and $\V$ be a set of views. 
If $r$ is returned by {\sf Count\_Rewriting}$(q,\V)$, then $r$ is a $\COUNT$-rewriting 
candidate of $q$ and $r \equivv q$.
If $q$ is either linear or relational, then the opposite holds
by making the appropriate choices.
\end{theorem}

\begin{example}\label{deductive:example}
This example shows the incompleteness of the algorithm if the 
comparisons in the query being rewritten are not deductively closed.
Consider the following query $q$, and views $v_1$ and $v_2$,
defined as 
\begin{eqnarray*}
 q(\COUNT) &\qif  &p_1(x)\AND p_2(y)\AND x<y \AND y<2  \AND{}\\ 
	& &	p_3(u)\AND p_4(w) \AND u<w \AND w<2 \\
v_1(x,u;\COUNT) &\qif & p_1(x)\AND p_2(y)  \AND x<y\AND y<2 \AND 
	 	p_3(u) \AND u<2  \\
v_2(x,u;\COUNT) &\qif & p_3(u)\AND  p_4(w)\AND u<w\AND w<2 
	 	 \AND p_1(x) \AND  x<2.   
\end{eqnarray*}

\noindent Note that the comparisons in $q$ are not deductively closed since 
$q$ does not contain $x<2$ and $u<2$. The algorithm {\sf Count\_Rewriting}
will not find any rewritings of $q$ using $v_1$ and $v_2$. We can
understand
this in the following way. Suppose that the view $v_1$ is chosen first.
Clearly, $v_1$ can be used for a rewriting taking the instantiation
$\theta$
and the isomorphism $\varphi$ to be the identity mappings. 
The algorithm {\sf Count\_Rewriting}
removes from $q$ the comparisons $x<y$ and $y<2$ since they contain the 
variable $y$ which is an image of the nondistinguished variable $y$ in
$v_1$.
However, {\sf Count\_Rewriting} can no longer use $v_2$ in the rewriting
since the constraints in $q$ no longer imply the constraint $x<2$ in $v_2$.
For symmetric reasons, {\sf Count\_Rewriting} would also fail to find a
rewriting
if $v_2$ was chosen first.
However, clearly the following is a rewriting
of $q$ using $v_1$ and $v_2$:
\begin{center}
$r(\SUM(z_1*z_2)) \qif  v_1(x,u;z_1)\AND v_2(x,u;z_2)$.  
\end{center}
\end{example}

\begin{example}
The algorithm is incomplete for the general case.
Consider the following query $q$, and view $v$ %, rewriting $r$, and unfolding $\unf r$
\begin{eqnarray*}
 q(;\COUNT) &\qif  &p(x)\AND p(y)\AND p(u)\AND x<y \AND x<u \\
 v(;\COUNT) &\qif  &p(x')\AND p(y')\AND p(u')\AND x'<y'\AND u'<y' 
\end{eqnarray*}

Clearly $q$ and $v$ are equivalent~\cite{Nutt:Et:Al-Equivalences:Among:Aggregate:Queries-PODS}.
However, for all homomorphisms $\varphi$ from $v$ to $q$, it holds that 
$\{x<y \AND x<u\} \not\models \{\varphi(x'<y') \AND \varphi(u'<y') \}$.
Thus, $v$ is not $C$-usable for rewriting $q$ and the algorithm does not find any rewritings. 
\end{example}

\subsection{Rewritings of Sum-Queries}
%%%%%%%%%%%%%%%%%%%%%%%%%%%%%%%%%%%%%%%%%%%%%%%%%%%%%%%%%%%%%%%%%%%%%%%%%%%%
Rewriting $\SUM$-queries is similar to rewriting $\COUNT$-queries. When 
rewriting $\SUM$-queries we must also take the summation variable into consideration.
We present an algorithm for rewriting $\SUM$-queries that is based on the
algorithm for $\COUNT$-queries. 

We define the form of rewriting candidates for $\SUM$-queries. Since 
$\SUM$ and $\COUNT$-views are the only views that are sensitive to multiplicities, 
they are useful for rewritings.
However, $\SUM$-views may lose multiplicities and make the aggregation variable inaccessible. 
Thus, at most one $\SUM$-view should be used in the rewriting of a query. The following
are rewriting candidates for $\SUM$-queries:
\begin{eqnarray}
&&\!\!\!\!r_1(\tpl x; \SUM(y * {\displaystyle\prod_{i=1}^n} z_i))
       \qif  %%%%\\ \nonumber
%%% &&\hspace{0.3in}
\vc_1(\theta_1 \tpl x_1; z_1) \AND \ldots \AND \vc_n(\theta_n \tpl x_n; z_n) \AND C' \label{pure:sum:cand}\\
&&\!\!\!\!r_2(\tpl x; \SUM(y * {\displaystyle\prod_{i=1}^n} z_i))   
        \qif \vs(\theta_\Sum \tpl x_\Sum; y) \AND %%%%%\\ \nonumber
%%%%% 	&&\hspace{0.3in}
\vc_1(\theta_1 \tpl x_1; z_1) \AND \ldots \AND \vc_n(\theta_n \tpl x_n; z_n) \AND C' \label{alpha:sum:cand}
\end{eqnarray}

\noindent where $\vc_i$ is a $\COUNT$-view of the form $\vc_i(\tpl x_i;\COUNT)\qif B_i$ and $\vs$
is a $\SUM$-view of the form
$\vs(\tpl x_\Sum;\SUM(y))\qif B_\Sum$. Note that the variable $y$ in the head of the
query in Equation~\ref{pure:sum:cand} must appear among $\theta_i \tpl x_i$ for some $i$.
In~\cite{Cohen:Et:Al-Rewriting:Aggregate:Queries-PODS}, it has been
shown that if a rewriting candidate is equivalent modulo the views to its unfolding 
then it must be one of the above forms.
As in the case of $\COUNT$-query rewritings, in some cases the rewriting
 may be optimized by dropping the summation.

Once again, we reduce reasoning about rewriting
candidates to reasoning about conjunctive aggregate queries. For this purpose we 
extend the unfolding technique introduced in Subsection~\ref{count:rewriting}.
Thus, the unfoldings of the candidates presented are:
\begin{eqnarray*}
        \unf r_1(\tpl x; \SUM(y)) &\qif&
                \theta_1 B_1 \AND \ldots \AND  
                \theta_n B_n \AND
                C'.\\
        \unf r_2(\tpl x; \SUM(y)) &\qif&
                        \theta_\Sum B_\Sum \AND
                        \theta_1 B_1 \AND \ldots \AND 
                        \theta_n B_n \AND
                        C'.
\end{eqnarray*}

Now, instead of checking whether $r$ is a rewriting of $q$ we can verify 
whether $\unf r$ is
equivalent to $r$. The only known algorithm for checking equivalence of 
$\SUM$-queries, presented in~\cite{Nutt:Et:Al-Equivalences:Among:Aggregate:Queries-PODS},
requires an exponential blowup of the queries. 
However, relational $\SUM$-queries and linear 
$\SUM$-queries are equivalent if and only if they are isomorphic. Thus, we can extend the 
algorithm presented in the 
Figure~\ref{fig:count:query:rew-algo} for $\SUM$-queries. 

\begin{figure*}[htb]
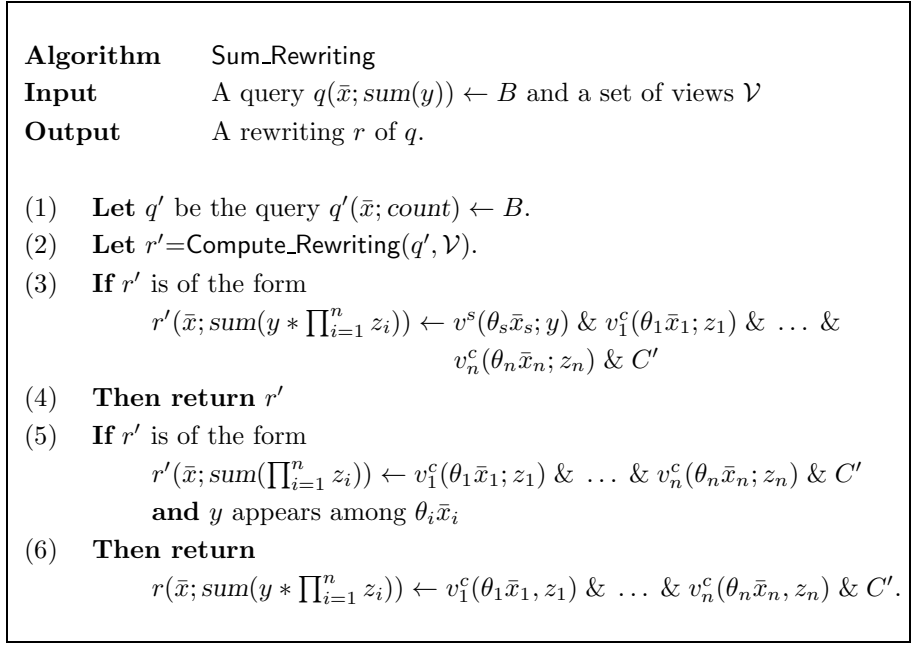

\def\baselinestretch{1.2}\small\normalsize
\begin{center}
\fbox{
\parbox{6in}{ %%%\tt
\begin{tabbing}
AAA \= AAA\= AAA\= AAAAA\= AAAAA\= AAAAA\= AAAAA\=\kill
{\bf Algorithm} \>\>\> {\sf Sum\_Rewriting}\\
{\bf Input}     \>\>\> A query $q(\tpl x;\SUM(y))\qif B$ and a set of views $\V$\\
{\bf Output}    \>\>\> A rewriting $r$ of $q$.\\
\\
(1) \> {\bf Let} $q'$ be the query $q'(\tpl x;\COUNT)\qif B$. \\
(2) \> {\bf Let} $r'$={\sf Compute\_Rewriting}$(q',\V)$. \\
(3) \> {\bf If} $r'$ is of the form \\
    \>	\> $r'(\tpl x; \SUM(y * \prod_{i=1}^n z_i))   
        			\qif
                \vs(\theta_\Sum \tpl x_\Sum; y) \AND
                \vc_1(\theta_1 \tpl x_1; z_1) \AND \ldots \AND$ \\
    \> \> \> \> \> \ \ \ \ \ $\vc_n(\theta_n \tpl x_n; z_n) \AND C'$ \\
(4) \> {\bf Then return} $r'$\\
(5) \>	{\bf If} $r'$ is of the form \\
	\>	\>$r'(\tpl x; \SUM(\prod_{i=1}^n z_i))
        \qif
                \vc_1(\theta_1 \tpl x_1; z_1) \AND \ldots \AND 
                \vc_n(\theta_n \tpl x_n; z_n) \AND C'$ \\
	\>	\> {\bf and} $y$ appears among $\theta_i\tpl x_i$  \\
(6) \>	{\bf Then return} \\
   \>	\>$r(\tpl x; \SUM(y * \prod_{i=1}^n z_i))
        \qif
                \vc_1(\theta_1 \tpl x_1, z_1) \AND \ldots \AND 
                \vc_n(\theta_n \tpl x_n, z_n) \AND C'$.
\end{tabbing}
}}
\caption{Sum Query Rewriting Algorithm}
\label{fig:sum:query}
\end{center}
\def\baselinestretch{2.0}\small\normalsize
\end{figure*}

%As a preliminary step 
We first extend the algorithm in 
Figure~\ref{fig:count:query:rew-algo}, such that in Line 5 $\SUM$-views may be chosen
as well. We call this algorithm {\sf Compute\_Rewriting}. We derive an algorithm 
for rewriting $\SUM$-queries, presented in Figure~\ref{fig:sum:query}. 
The algorithm runs in nondeterministic
polynomial time.

\begin{theorem}{\bf (Soundness and Completeness of Sum Rewriting)}
Let $q$ be a  $\SUM$-query and $\V$ be a set of views.
If $r$ is returned by {\sf Sum\_Rewriting}$(q,\V)$, then $r$ is a $\SUM$-rewriting 
candidate of $q$ and $r \equivv q$.
If $q$ is linear or relational, then the opposite holds
by making the appropriate choices.
\end{theorem}

\subsection{Rewritings of Max-Queries}\label{ssec-max:queries}
We consider the problem of rewriting $\MAX$-queries. 
Note that  $\MAX$-queries are insensitive to multiplicities. Thus, we
use nonaggregate views and $\MAX$-views when rewriting a $\MAX$-query. When
using a $\MAX$-view the aggregation variable becomes inaccessible.
Thus, we use at most one $\MAX$-view. The following
are rewriting candidates of the query $q$:
\begin{eqnarray}
r_1(\tpl x; \MAX(y))
      & \qif &
v_1(\theta_1 \tpl x_1) \AND \ldots \AND v_n(\theta_n \tpl x_n) \AND C' \label{pure:max:cand}\\
r_2(\tpl x; \MAX(y))   
        &\qif &\vm(\theta_\Max \tpl x_\Max; y) \AND
v_1(\theta_1 \tpl x_1) \AND \ldots \AND v_n(\theta_n \tpl x_n) \AND C' \label{alpha:max:cand}
\end{eqnarray}

Note that the $v_i$'s are nonaggregate views and that $\vm$
is a $\MAX$-view. The variable $y$ in the head of the
query in Equation~\ref{pure:max:cand} must appear among $\theta_i \tpl x_i$ for some $i$.
In~\cite{Cohen:Et:Al-Rewriting:Aggregate:Queries-PODS} it has
been shown that if a rewriting candidate is equivalent to its unfolding then 
it must have one of the above forms. 

Reasoning about rewriting candidates can be reduced
to reasoning about $\MAX$-queries, by extending the unfolding technique.
It has been shown~\cite{Nutt:Et:Al-Equivalences:Among:Aggregate:Queries-PODS} that 
equivalence of relational $\MAX$-queries is equivalence of their cores. There
is a similar reduction for the general case. Algorithms for checking 
set-equivalence of queries can easily be converted to algorithms for checking 
equivalence of $\MAX$-queries. Thus, algorithms that find rewritings of 
nonaggregate queries can be modified to find rewritings of $\MAX$-queries.

Rewriting nonaggregate queries is a well known 
problem~\cite{Levy:Et:Al-Reusing:Views-PODS}. 
%Thus, we do not present algorithms for finding rewritings of $\MAX$-queries.
One well-known algorithm for computing rewritings of queries is the
{\em buckets algorithm\/} \cite{LRO96,LRO96:VLDB}. Consider
a query $q(\tpl x)\qif R \AND C$. The algorithm creates a ``bucket'' for
each atom $p(\tpl z)$ in $R$. Intuitively, this bucket contains all the
views whose bodies can cover $p(\tpl z)$. The algorithm places into this bucket 
 all the views 
$v(\tpl y)\qif R_v \AND C_v$ such that $R_v$ contains an atom $p(\tpl w)$ 
that can 
be mapped by some mapping $\varphi$ to $p(\tpl z)$ such that $C\AND
\varphi C'$ 
is consistent. Next, all combinations of taking a view from each bucket are 
considered in the attempt to form a rewriting. 

Note that by reasoning similarly as in the case of $\COUNT$ and $\SUM$-queries,
we can reduce the number of views put into each bucket,
thus improving on the performance of the algorithm.
Suppose there is a nondistinguished 
variable $w\in \tpl w$ mapped to $z\in\tpl z$ and there is an atom 
containing $z$ in $q$ that is not covered by $\varphi R_v$. In such a case,
if $v$ is used in a rewriting candidate %to cover $p(\tpl z)$
there will not exist a homomorphism from the unfolded rewriting to $q$
such that the body of $v$ covers $p(\tpl z)$.
However, a rewriting candidate $r$ is equivalent to a query $q$ if and only 
if there exist homomorphisms from $\unf r$ to $q$ and from $q$ to $\unf r$.
Thus, $v$ should not be put in the bucket of $p(\tpl z)$. 

Observe that 
this condition is a relaxed version of the $R$-usability requirement that 
ensures the existence of an isomorphism.
Clearly this restriction filters out the possible rewritings of $q$, 
thereby improving the performance of the buckets algorithm.
Thus, our methods for finding rewritings of aggregate queries may be relaxed
and used to improve the performance of algorithms for rewriting relational
queries. These, in turn, may be modified to rewrite $\MAX$-queries.

%% file: conclusion.tex
\section{Conclusion}
%%%%%%%%%%%%%%%%%%%%%%%%%%%%%%%%%%%%%%%%%%%%%%%%%%%%%%%%%%%%%%%%%%%%%%%%%%%%%%%%
Aggregate queries are increasingly prevalent due to the widespread use of 
data warehousing and related applications. They are generally computationally
expensive since they scan many data items, while returning few results.
Thus, the computation time of
aggregate queries is generally orders of magnitude larger than the result
size of the query. This makes query optimization a necessity. 

Optimizing aggregate queries using views has been studied for the special
case of
datacubes~\cite{Harinarayan:Et:Al-Efficient:Data:Cubes-SIGMOD,%
        Dyreson-Incomplete:Datacube-VLDB}.
However, 
there was little theory for general aggregate queries.
In this paper, based on previous results 
in~\cite{Nutt:Et:Al-Equivalences:Among:Aggregate:Queries-PODS,%
	Cohen:Et:Al-Rewriting:Aggregate:Queries-PODS},
 we presented algorithms that enable reuse of precomputed 
queries in answering new ones. 
The algorithms presented have been implemented in SICStus 
Prolog.

\eat{ at Hebrew University. The system is located at  
{\tt \verb+http://droopy.cs.kuleuven.ac.be:8000/+}
{\tt \verb+alexander/AggrQ/main_menu.html+}.%www.cs.huji.ac.il/+}
{\tt \verb+~alicser/aggrq/+}. }

\eat{
Topics for future research include rewriting queries with {\tt HAVING} clauses,
negation and functional dependencies, and enriching the class of aggregate
 functions with statistical functions. 
}

%% file: appendix.tex
\appendix

\section{Translating SQL to Datalog}
	\label{appendix}
%%%%%%%%%%%%%%%%%%%%%%%%%%%%%%%%%%%%%%%%%%%%%%%%%%%%%%%%%%%%%%%%%%%%%%%%

In this paper we extended the well-known Datalog syntax for non-aggregate
queries~\cite{Ullman-Database:And:K:B:Systems:II}
so that it covers also aggregates.
This syntax is more abstract and concise than SQL.
It is not only better suited for a theoretical investigation, 
but it is also a better basis for implementing algorithms that reason
about queries, in particular for implementations in a logic
programming language.

Through the syntax we implicitly define the set of SQL queries to
which our techniques apply.  They are essentially nonnested queries
without a {\tt HAVING} clause and with the aggregate operators $\MIN$, $\MAX$,
$\COUNT$, and $\SUM$. In this section we demonstrate, using examples,
 how an SQL query of this
type can be transformed into one in our extended Datalog notation.

We first show how to transform an SQL query without aggregation into
one in Datalog notation. Consider a query with {\tt SELECT}, {\tt FROM},
and {\tt WHERE} clauses. For each relation name in the {\tt FROM} clause
we introduce a predicate name,  and
for each attribute of a relation, we fix an argument position of the
corresponding predicate. For each occurrence of a relation name in the 
{\tt FROM} clause we create a relational atom.
The selection constraints in the {\tt WHERE} clause are taken into
account by placing constants or identical variables into
appropriate argument positions of the atoms corresponding to a relation, 
or by imposing comparisons on variables.
Finally, the output arguments in the {\tt SELECT} clause appear as the
distinguished variables in the head.%%% of $\sfq$.

We demonstrate the translation
using an example. This example can easily be generalized to arbitrary
SQL queries without  {\tt GROUP-BY} and {\tt HAVING} clauses.
Consider a query that finds the teaching assistants who have a job for
which they receive more then \$500 from the government:

\noindent
\mbox{}\qquad%
\parbox[t]{30em}{%
\tt
\begin{tabbing}
AAAAAAAA\=A\kill
SELECT     \> {\sf name}\\
FROM       \> {\sf ta, salaries}\\
WHERE      \> {\sf sponsorship} = 'Govt.' AND  {\sf amount} $>$ 500\\
           \> AND {\sf ta.job\_type} = {\sf salaries.job\_type}{\rm .}
\end{tabbing}}

We translate this query into an equivalent Datalog query with the
head predicate $\tt q\_govt$.
For the relation names {\sf ta} and {\sf salaries} we introduce
the predicate names {\tt ta} and {\tt salaries}. In the fashion 
described above, we derive the following equivalent Datalog query:
\begin{eqnarray*} \label{ex-non:agg:datalog}
{\tt q\_govt}(n) & \qif &  {\tt salaries}(j,{\tt Govt.},a) \AND
			{\tt ta}(n,c,j) \AND 
		  a > 500.
\end{eqnarray*}

In this paper we extended the Datalog syntax so as to capture also queries with
{\tt GROUP BY} and aggregation.
Using our notation, we can represent SQL queries where the group by 
attributes are {\em identical\/} to those in the {\tt SELECT} statement,
although SQL only requires that the latter be a {\em subset\/}
of those appearing in the {\tt GROUP BY} clause.
Also, we assume that queries have only one aggregate term.
The general case can easily be reduced to this one.

The extension of the Datalog syntax is straightforward.
Since the {\tt SELECT} attributes are identical to the grouping
attributes, there is no need to single them out by a special notation.
Hence, the only new feature is the aggregate term in the {\tt SELECT}
clause.
We simply add it to the terms in the head of the query, after
replacing the attributes with corresponding variables.

To demonstrate this translation, recall the query in 
Section~\ref{sec-motivation} that calculates
the total amount of money spent on each job type. 
The following SQL query can be transformed into the previously mentioned 
Datalog query:

\noindent
\mbox{}\qquad%
\parbox[t]{30em}{%
\tt
\begin{tabbing}
AAAAAAAA\=A\kill
SELECT     \> {\sf ta.job\_type}, \SUM({\sf amount})\\
FROM       \> {\sf ta, salaries}\\
WHERE      \> {\sf ta.job\_type = salaries.job\_type}{\rm .}
\end{tabbing}}

We have demonstrated how to translate SQL into Datalog. Obviously,
the translation from Datalog to SQL can be performed in a similar fashion.
Roughly speaking, we replace predicates with relations and 
variables with attributes. The variables in the head of the Datalog query
become attributes in the {\tt SELECT} clause of the SQL query and
the comparisons are placed in the {\tt WHERE} clause.
Hence, one notation can be transformed into the other, back and
forth, completely automatically.

%% file: adbis.bbl
\begin{thebibliography}{LRO96b}

\bibitem[BI94]{BI94}
D.~Barbara and T.~Imielinski.
\newblock Sleepers and workaholics: Caching strategies in mobile environments.
\newblock 1994.

\bibitem[CNS99]{Cohen:Et:Al-Rewriting:Aggregate:Queries-PODS}
S.~Cohen, W.~Nutt, and A.~Serebrenik.
\newblock Rewriting aggregate queries using views.
\newblock 1999.
\newblock To appear.

\bibitem[CV93]{Chaudhuri:Vardi-Real:Conjunctive:Queries-PODS}
S.~Chaudhuri and M.~Vardi.
\newblock Optimization of real conjunctive queries.
\newblock 1993.

\bibitem[Dyr96]{Dyreson-Incomplete:Datacube-VLDB}
C.~Dyreson.
\newblock Information retrieval from an incomplete datacube.
\newblock 1996.

\bibitem[GHQ95]{Gupta:Et:Al-Generalized:Projections-VLDB}
A.~Gupta, V.~Harinarayan, and D.~Quass.
\newblock Aggregate query processing in data warehouses.
\newblock 1995.

\bibitem[HRU96]{Harinarayan:Et:Al-Efficient:Data:Cubes-SIGMOD}
V.~Harinarayan, A.~Rajaraman, and J.~Ullman.
\newblock Implementing data cubes efficiently.
\newblock pages 205--227, 1996.

\bibitem[Klu88]{Klug-Conjunctive:Queries:Inequalities-JACM}
A.~Klug.
\newblock On conjunctive queries containing inequalities.
\newblock {\em Journal of the ACM}, 35(1):146--160, 1988.

\bibitem[LMSS95]{Levy:Et:Al-Reusing:Views-PODS}
A.Y. Levy, A.O. Mendelzon, Y.~Sagiv, and D.~Srivastava.
\newblock Answering queries using views.
\newblock pages 95--104, 1995.

\bibitem[LRO96a]{LRO96:VLDB}
Alon Levy, Anand Rajamaran, and Joann Ordille.
\newblock Querying heterogeneous information sources using source description.
\newblock In {\em Proceedings of the 22nd VLDB Conference Mumbai(Bombay),
  India}, 1996.

\bibitem[LRO96b]{LRO96}
A.Y. Levy, A.~Rajaraman, and J.J. Ordille.
\newblock Query answering algorithms for information agents.
\newblock In {\em Thirteenth National Conf. on Artificial Intelligence,
  AAAI-96}, 1996.

\bibitem[LSK95]{Levy:Et:Al-Global:Information:Systems-JIIS}
A.Y. Levy, D.~Srivastava, and T.~Kirk.
\newblock Data model and query evaluation in global information systems.
\newblock 5(2):121--143, 1995.

\bibitem[NSS98]{Nutt:Et:Al-Equivalences:Among:Aggregate:Queries-PODS}
W.~Nutt, Y.~Sagiv, and S.~Shurin.
\newblock Deciding equivalences among aggregate queries.
\newblock pages 214--223, 1998.
\newblock Long version as Report of Esprit LTR DWQ.

\bibitem[Qia96]{Qian-Query:Folding-ICDE}
X.~Qian.
\newblock Query folding.
\newblock pages 48--55, 1996.

\bibitem[SDJL96]{Srivastava:Et:Al-Reusing:Views:With:Aggregates-VLDB}
D.~Srivastava, Sh. Dar, H.V. Jagadish, and A.Y. Levy.
\newblock Answering queries with aggregation using views.
\newblock 1996.

\bibitem[TS97]{Theodoratos:Sellis-D:W:Configuration-VLDB}
D.~Theodoratos and T.K. Sellis.
\newblock Data warehouse configuration.
\newblock pages 126--135, 1997.

\bibitem[Ull89]{Ullman-Database:And:K:B:Systems:II}
J.~Ullman.
\newblock {\em Principles of Database and Knowledge-Base Systems, Vol. II: The
  New Technologies}.
\newblock Computer Science Press, New York (New York, USA), 1989.

\end{thebibliography}
